# High-Throughput Atomic Force Microscopes Operating in Parallel

H. Sadeghian, R. Herfst, B. Dekker, J. Winters, T. Bijnagte, and R. Rijnbeek

Department of Optomechatronics, Netherlands Organization for Applied Scientific Research, TNO, Delft, The Netherlands

**Abstract**

Atomic force microscopy (AFM) is an essential nanoinstrument technique for several applications such as cell biology and nanoelectronics metrology and inspection. The need for statistically significant sample sizes means that data collection can be an extremely lengthy process in AFM. The use of a single AFM instrument is known for its very low speed and not being suitable for scanning large areas, resulting in very-low-throughput measurement. We address this challenge by parallelizing AFM instruments. The parallelization is achieved by miniaturizing the AFM instrument and operating many of them simultaneously. This nanoinstrument has the advantages that each miniaturized AFM can be operated independently and that the advances in the field of AFM, both in terms of speed and imaging modalities, can be implemented more easily. Moreover, a parallel AFM instrument also allows one to measure several physical parameters simultaneously; while one instrument measures nano-scale topography, another instrument can measure mechanical, electrical or thermal properties, making it a Lab-on-an-Instrument. In this paper, a proof of principle (PoP) of such a parallel AFM instrument has been demonstrated by analyzing the topography of large samples such as semiconductor wafers. This nanoinstrument provides new research opportunities in the nanometrology of wafers and nanolithography masks by enabling real die-to-die and wafer-level measurements and in cell biology by measuring the nano-scale properties of a large number of cells.

**I. INTRODUCTION**

Atomic force microscopy (AFM) instruments are an essential type of nanoinstrument and have contributed to major breakthroughs in materials research,[1] nanoelectronics[2] and biology.[3] Today, AFM is used for not only sub-nanometer imaging but also manipulation and manufacturing at the nano-scale.[4] Using various modes of tip-sample interactions, such as atomic force,[5] near-field optics,[6] electrostatic,[7] thermal[8] and electromagnetic forces,[9] AFM instruments have evolved toward becoming a Lab-on-an-Instrument.

Traditional AFM uses a singular head/cantilever entity. This poses a limit on the throughput of imaging, characterization and nanomanufacturing. It also limits true sample-to-sample measurement comparison, which is beneficial, for example, in distinguishing normal cells from abnormal or cancerous cells.[10] The increase in the imaging speed in AFM is of interest in several nanotechnology applications, including wafer and photolithography mask metrology in nanoelectronics applications[10,11] and investigations of living cells[12] and proteins[3] in both biology and medicine. Single AFM has never been able to compete with other inspection systems in terms of throughput and thus has not fulfilled industry needs in that respect. Further increases in the speed of single AFM helps, but it still is far from the required throughput and, therefore, insufficient for applications that require statistically significant sample sizes or large area, high-resolution measurements such as of wafers and photolithography masks. In these applications, data collection can be an extremely lengthy process.

Several research groups have recently been enhancing the speed of AFM for the imaging of dynamic processes, mostly in the field of biophysics, with impressive results.[13-16] In these cases, the sample size is usually very small, e.g., a few mm, and the sample has small height corrugations. This makes "sample scanning" design a suitable architecture, in which the AFM cantilever is kept stationary and the sample is scanned along three axes. The optical read-out system and the clamping mechanism of the cantilever are easier to implement in this way. Due to the small scan size and low weight of the sample, a very high resonance frequency of the sample stage can be achieved, enabling a very high scanning bandwidth, which is essential for high-speed AFM.[13] This architecture, however, is not suitable for large samples such as wafers, photolithography masks and arrays of bio-samples such as organs. Because such samples have substantial masses, the resonance frequency of the sample stage, especially in the vertical direction, becomes extremely low, resulting in low measurement speeds.

Several research groups have suggested parallelization of AFM as a direct approach to increasing AFM throughput in direct proportion to the number of AFM tips in an array.[12,14-18] Following this path, efforts to increase the throughput of AFM using arrays of cantilevers have been made, instead of a single cantilever, motivated by advances in microsystems fabrication technologies and CMOS processes. Examples include data storage,[18] parallel imaging and force spectroscopy[12] and parallel AFM nanolithography.[17] There remain several significant challenges in developing a massively parallel AFM instrument to be used in practical applications. Examples of remaining challenges include signal processing, probe positioning, the limited number of wires, the exchange of one probe in an array, cross talk, data transfer, and the reliability of the system.

Instead of an array of microfabricated tips, we opted for a second level at which AFM can be parallelized: miniaturizing the AFM instrument and operating many of them in parallel.[11] This architecture has the advantages that each miniaturized AFM (MAFM) instrument can be both operated and positioned independently. Moreover, when desired, an array of cantilevers can still be used as an additional level of parallelization within each MAFM instrument.[19] We have achieved a very high throughput in AFM via three developments:

1) A speed increase in MAFM by increasing the bandwidth of the AFM's sub-modules, i.e., the mechanical stages (x, y and z), optical read-out,[20,21] controller bandwidth, approach speed and speed of positioning.[22] We have experimentally demonstrated such a high-speed MAFM instrument[19] and a mini positioning unit (PU) capable of positioning each MAFM instrument very quickly and accurately.[23]

2) Sufficient miniaturization of the AFM instrument to operate many such instruments in parallel.

3) Development of a system architecture that can handle relatively large samples.

This paper presents a proof of principle (PoP) and experimental demonstration of the parallel AFM system, where a number of MAFMs scan several sites on a semiconductor wafer. This proves the feasibility of such a parallel AFM system for high-throughput nano-scale measurements in several applications. Moreover, several physical parameters of samples can be measured simultaneously in parallel, which enables true sample-to-sample comparison. For this, each MAFM instrument monitors a specific tip-sample interaction such as atomic, thermal, electrical or magnetic forces.

This paper is organized as follows: Section 1 describes the overall architecture of the parallel AFM instrument and the performance specifications. In Section 2, the design and realization of the parallel AFM instrument and sub-modules, such as the MAFM instrument, miniaturized PU and large sample stage, are discussed. Section 3 presents the experimental results obtained with the system.

**II. ARCHITECTURE OF PARALLEL AFM**

The core of the architecture of the parallel AFM is based on an MAFM instrument and a miniature PU to position the MAFM instrument at the targeted scanning location. The objective is to parallelize many such MAFM instruments and PUs to be able to scan many sites in parallel. Figure 1 shows the concept of such a parallel AFM instrument.[24] There are several options for implementing parallel AFM systems. We opted for a design in which the AFM instruments are simplified as much as possible, with their main functionality being achieving a high vertical bandwidth of measurement with high vertical and lateral resolutions. Each AFM instrument should be miniaturized as much as possible to maximize the number of instruments that can be operated in parallel. All the MAFM instruments are stationary relative to each other in the x,y- directions.

The sample (in this case, a silicon wafer) will be on a customized stage that facilitates lateral x,y- scanning. In this way, all MAFM instruments will perform their scans in a similar pattern, each at its particular site on the wafer. This enables a fully decoupled lateral scanner from the z stage that moves the probe. This eliminates the well-known scanning bowing issue (i.e., obtaining a very flat x,y- scan) and substantially increases the response of the z-scanning stage.

We opted for a maximum image size of 25×25 $\mu m^2$, which determines the stroke of the sample stage. The resolution of the sample stage is chosen to be 0.1 nm over its entire bandwidth. The sample stage movement is controlled in a closed loop using capacitive sensors.

Completely fixed x,y-positions of the MAFM instruments would greatly limit the flexibility. Therefore, the MAFM instruments can individually be positioned relative to the wafer prior to the scanning of the wafer. To keep the design simple, each MAFM instrument can only be positioned on a part (strip) of the wafer by an arm with a large stroke in one direction (x) and a small stroke in the other direction (y). In this way, a set of parallel MAFM instruments can cover the full wafer, and they can be moved onto and off of the wafer to enable loading and unloading of the wafer.

The maximum width of each MAFM instrument (including arm) is 19 mm, enabling the operation of 44 parallel MAFM instruments for a 450 mm wafer (30 for a 300 mm wafer), 22 (15 for a 300 mm wafer) for each side of the wafer. The PoP described in this paper consists of 4 parallel PUs, each containing an MAFM instrument that simultaneously scans a wafer or a mask. The user can define several locations to be scanned. Figure 2 shows the final design of the parallel AFM system.

To successfully integrate multiple PUs and MAFMs into a full system, each PU and MAFM instrument combination needs to be contained within a narrowly defined area. In the first generation of MAFM instruments and PUs,[23] this could not be realized because all the electrical components were connected with individual wires and cables. Therefore, we have integrated the required electrical connections into flex-rigid PCBs that are glued to the mechanical body of the PU. The demonstration of the PoP is given in Figure 3. The wafer stage is shown in the middle. The PUs and MAFM instrument are located at the two sides of the system.

## III. DETAILED DESIGN OF THE MODULES

In this section, the design of the key modules, i.e., the PU and PU metrology, MAFM instrument and wafer stage, is discussed.

### A. Positioning unit and positioning metrology

The PU has the primary function of positioning a MAFM instrument in the horizontal plane with respect to the scanned sample. The PU consists of a mechanical arm and associated actuators and sensors to perform the positioning function. This is schematically shown in Figure 4.

The arm is mounted onto a linear motor with encoder, which is mounted on a plate. This plate represents the fixed part of the PU and is stiffened by the linear motor assembly. The linear motor enables positioning of the arm, including the MAFM instrument, in the x-direction. The arm has a built-in flexure-based mechanism that makes it possible to position the MAFM instrument in the y-direction.[25] When the y-mechanism is in the nominal position, the width of the entire PU assembly, including cabling, fits in the required 19-mm-wide envelope.

The position of the cantilever tip with respect to the wafer and targeted scanned area must be known before and during the AFM measurements. The positioning loop and metrology loop are illustrated in Figure 4. To obtain the required positioning accuracy for the tip, i.e., 1 µm in the lateral direction, closed-loop control is implemented in the PU. Interferometry cannot be used to measure the positions of the PUs because the parallel PUs obstruct each other's view in the x- and y-directions. Instead, an x,y-grid-based optical linear encoder was chosen for this metrology. We use three sensor heads in the base of each MAFM instrument: one in the x-direction to measure the long axis of the movement and two in the y-direction to measure the other axis as well as the in-plane rotation. The rotation information is required to decouple translation in the y-direction and rotation and, thus, accurately determine the tip position. The PU consists of a long and narrow mechanical arm, which has a low eigenfrequency of 80 Hz.[19] The connection of this low stiff arm to the MAFM instrument during scanning will induce noise in the measurement. Therefore, the PU needs to be decoupled from the MAFM instrument because the vibration amplitude at the PU due to the floor and other environmental excitations is excessive. The decoupling of the PU from the MAFM instrument during scanning is realized using the MAFM coarse approach system. First, the bottom of the MAFM instrument is lowered onto a stiff frame. Once the MAFM instrument has landed on the frame, the top side approaches the sample using the same approach motor. Consequently, the PU is no longer attached to the MAFM instrument and is out of the mechanical loop for the measurement. This is also shown in Figure 4. As seen, the measurement loop is considerably shorter than the positioning loop, both depicted in Figure 4. Capacitive sensors in the wafer stage provide real-time feedback of the wafer stage position. A large grid plate is placed directly on the stone and provides the location reference for the MAFM instruments.

Figure 5 shows the PoP of the PU and the MAFM instrument. A total of 22 PUs can be located in the 22 slots as follows. The PUs can be placed at discrete positions beside each other with a pitch of 20 mm. We designed a cassette box with 22 slots to hold the PUs in place, as shown in Figure 5. The PUs are manually loaded into the cassette box from the top. Any number of PUs in any order can be chosen in these 22 slots for maximum flexibility. The PUs in the cassette box can reach from the outside edge of a wafer to the wafer center. A second cassette box is added to enable scanning of the opposing half of the wafer, thus increasing the maximum number of scan sites to 44.

The PU's z-position and y-rotation with respect to the grid plate can be adjusted with two adjustment screws.

In total, 69 electrical connections were needed in the PU arm. Given the available design volume, this was not feasible with normal wires. Instead, the connections were made with multiple flex-rigid cables. Five of these cables cross over from the MAFM instrument to the PU arm and are folded such that each flex-rigid crossover has a planar section in each of the three possible orthogonal planes; see Figure 6. This dynamic bridge sufficiently reduces the cable stiffness in all six degrees of freedom, thus minimizing disturbance forces coming from the arm that would compromise the stability of the positioned MAFM instrument.

The distribution of the wires in multiple groups made it possible to simultaneously isolate critical signal paths from noisy wires such as those carrying high voltage for driving the piezos. The location of all wires was optimized within each flex-rigid cable and also for the routing of all flex-rigid cables though the PU toward the fixed world.

**B. Miniature AFM**

The core development of the parallel AFM instrument relies on the development of a MAFM instrument. We recently developed such an MAFM instrument with a size of approximately 70×19×45 mm$^3$ to be implemented in the parallel AFM instrument.[19,22] A schematic illustration of the MAFM instrument is shown in Figure 7. The MAFM instrument has the following proven performance specifications. The flexure-based and counter mass balanced vertical z stage, which is used to adjust the distance between the sample and the probe, has a first eigenfrequency of 50 kHz. The MAFM frame has an eigenfrequency of 1 kHz, which is much lower than that of the vertical z stage.

The dynamics of the vertical z scanning stage are made insensitive to the surrounding parasitic forces by suspending the z scanning stage at specific dynamically determined points.[22] The dynamically determined points are the locations where the deformation of the stage is zero. They referred to as stationary points because only rotation (but not translation) can occur at these points. This approach minimizes the coupling of the motion with the frame. Further details have been reported elsewhere.[22]

Figure 8 shows the frequency response improvement from 1 kHz to 50 kHz when the z stage is suspended via dynamically determined points. The z stage has a net vertical travel range of 2.1 μm.

A compact optical beam deflection (OBD) readout has been realized to measure the motion of the cantilever. It has a resolution of 15 fm/√Hz and a bandwidth of 3 MHz, which allows the use of ultra-high frequency cantilevers.[19] A fast-approach mechanism positions the vertical z stage and OBD toward the sample. As explained earlier, the approach motor is used for the functions of landing the MAFM instrument on the frame, detaching from the PU, and approaching the sample. A contact transfer mechanism has been used to ensure that the MAFM instrument is detached from the arm during scanning and attached during retracting in a repeatable manner. Detachment and attachment are sensed using three contact switches implemented between the MAFM instrument and the PU. Three optical encoders, as shown in Figure 7, are used to measure the position of the MAFM instrument. All wires from the MAFM instrument are routed toward the rear of the MAFM instrument where they cross over to the arm. Figure 9 shows the realized MAFMs in parallel. In this figure, the second MAFM can also be seen.

The optical components of the OBD, i.e., the laser diode and the lens, are mounted into an optics barrel. An air gap between the diode and housing results in insufficient heat dissipation of the diode, making the lifetime of the diode very short. We have overcome this problem by press-fitting the diode into the barrel and by clamping the barrel to the MAFM instrument frame through a relatively large contact area. Consequently, the thermal resistance from the diode to the barrel housing is lowered, which is the most dominant resistance in the thermal path. Analysis showed that the maximum temperature rise in the new barrel design is 20 degrees.

The laser barrels were tested for potential overheating problems by operating them at the full rated power for seven hours; none of the barrels failed. Furthermore, one laser barrel was tested for a full day at 120% power without any failure. In the MAFM instrument, the lasers are operated at only 24% of the rated power, thereby further eliminating the risk of temperature-related failure of the laser diodes.

A fast-approach mechanism was also implemented in the MAFM instrument.[19] The sample is approached by continuously stepping closer to the surface while the AFM control loop is closed and tuned for a very fast response. Although it is possible to approach the sample while keeping the AFM tip intact, this requires a properly tuned fast feedback, which is only possible with high-bandwidth cantilevers.

To assure no damage to the tip or to the sample, the following steps are taken. The approach is divided into two phases: an initial fast phase where the coarse approach motor continuously moves the cantilever closer to the surface and a slower 'walk&talk' phase. During the latter approach, steps of approximately 1 µm ('walk') are interleaved with a sensing step ('talk'), during which the fine z-stage extends to check whether the sample can be reached and retracted again before the next coarse approach steps. The systems switches from the fast phase to the walk&talk phase when the amplitude starts to decrease due to the increase in squeeze film damping as the cantilever almost reaches the surface. For the data presented here, the trigger for switching was set at an amplitude reduction of 20%. Because only a few walk&talk cycles are required during the second phase, this can be done in a very safe manner (i.e., relatively slow movement of the fine z-stage when it is moving toward the surface). The results of the fast approach are shown in the experimental results section.

**C. Wafer scanning stage**

This section describes the development of the wafer x,y- scanning stage. The wafer stage provides the required sample motion for scanning in the x- and y-directions. The travel range is 50 µm in the x- and y-directions. For scanning an area of 10 × 10 µm$^2$ at a 5 nm grid density, the speed of the wafer stage was designed to be 12.5 Hz. The required noise levels are 0.1 nm (3σ) in-plane and 0.05 nm (3σ) out-of-plane.

The wafer-stage concept architecture includes a horizontally moving stage body consisting of a vacuum chuck for a number of wafer sizes. The moving stage is suspended in the vertical direction by a set of elastic hinges, which allow freedom of motion for x- and y-translations and for rotation about the z-axis. A separate mechanism on top of the moving stage provides the rotation constraint about the z-axis. The chuck is driven in the x- and y-directions by piezo stacks, which operate independently for each axis. Position feedback is provided by two capacitive sensors, one for the x-direction and one for the y-direction. A two-channel position controller was used to control the motions in the x- and y-directions. The controller was implemented in PFGA.

For control of the 12.5 lines/s scan motion, we specified a control bandwidth of at least 80 Hz. Therefore, we specified the lowest mechanical eigenfrequency at 300 Hz or higher. Requirements for the stage suspension are a vertical guide stiffness of 1e8 N/m or more in the x- and y-directions and 1.2e6 Nm/rad or more in rotation about the z-axis. In the x- and y-directions, the actuation stiffness requirement is 1e8 N/m or more, and the guide stiffness requirement is 1e5 N/m or less.

The detailed design of the wafer stage is depicted in Figure 10. A modal analysis of the designed wafer stage shows that the lowest eigenfrequency is 338 Hz and that the mode is the vertical displacement of the moving body. Therefore, the lowest eigenfrequency is well above the requirement of 300 Hz.

The realization of the wafer stage is shown in Figure 11. Experimental modal analysis showed a wafer stage moving body eigenfrequency of 292 Hz, which is lower than the modal analysis model but remains a factor of 3 above the 80 Hz controller bandwidth.

The piezo actuators were selected based on the required 50 µm total stroke. An effective maximal stroke of 33 µm was realized. The experimental results of the motion speed showed a line rate of 20 Hz at a 10.8 µm stroke; see Figure 11.

## IV. EXPERIMENTAL RESULTS

### A. Fast and safe approach

Figure 12 shows the measurement results of a fast approach for a sample from a distance of 0.5 mm. As seen, the amplitude does not change significantly for the initial 2 seconds. Immediately before t = 2.18 seconds, the amplitude starts to decrease, triggering the system to switch to the 'walk&talk' phase. For the threshold for switching that was chosen here (20% reduction), only one walk&talk cycle was necessary. After only 2.64 seconds, the system is fully engaged and ready to start imaging. The measured amplitude signal showed that at no point in time was the cantilever vibration fully quenched, indicating that the system did not overshoot while approaching the sample, which could damage the tip and/or the sample.

**B. AFM imaging experiments**

The experimental results of the AFM imaging on different samples are shown here. One of the applications of the parallel AFM is for nanoparticle detection. The objective is to detect particles or defects of 10 nm on a large area of 10 µm. The sample was made using mica substrates, with colloidal gold nanoparticles deposited on the substrates using the vendor's instructions.[26] Figure 13 (a) shows a 10x10 µm scan size of the colloidal gold nanoparticles on mica, with 2048x2048 pixels at a line-rate of 13 Hz. In the selected area within the solid square (2x2 µm), we clearly observe 10 different gold particles, although they were enlarged to 20–30 nm by the poly-L-lysine used to fixate the particles to the sample (the height was still 10 nm). Re-measuring this 2x2 µm area confirms that no particles have been missed. It also yields higher resolution details of the particles, showing the system's viability for particle and defect detection and review.

To demonstrate simultaneous imaging with four parallel MAFMs, four grating samples[27] were imaged. Figure 14 shows simultaneously scanned 10×10 µm$^2$ images of four calibration samples with 2048×2048 pixels.

The parallel AFM system uses a scanning wafer stage along with feedback to improve the scanning trajectory. We observed that, at line-rates of 10 Hz and above, some distortions are visible, especially at the edges where the stage is at its turnaround point. Since the x- and y-positions of the stage are measured with high-resolution capacitive sensors for feedback, we can correct this distortion. This is done by recording the lateral position values along with z- and error-signal data and using interpolation to map the z- and error-signal data to a square grid.

To assess the stability of the parallel AFM instrument against thermal drift and air flow, we measured a 3-µm-pitch, 110-nm-high grating line sample twice at an intentionally low speed of 2.5 Hz at 2048x2048 pixels. The total measurement time was approximately 27 minutes. The result of these two consecutive scans is shown in Figure 15. As seen, the lines are almost perfectly straight, with only a slight difference in angle and lateral offset between the two scans.

To calculate the average drift, we compared a cross-section from the start of the test to a cross-section at the same line at the end of the tests. This is shown in Figure 16. A total shift of 122 nm over a total time of 27 minutes was determined. This leads to an average drift of only 0.075 nm/s. The high stability of the parallel AFM demonstrator makes it a suitable metrology platform for applications such as critical dimension (CD) metrology, overlay measurement and photolithography mask metrology at the device level.

**V. CONCLUSIONS**

A parallel AFM demonstrator has been developed for high-throughput, sub-nanometer imaging, characterization and metrology of large area samples such as wafers and photomasks. The PoP demonstrator consists of several miniaturized AFM instruments and PUs, which enable many simultaneous measurement locations on a sample. Each PU can position a miniaturized AFM instrument in less than 5 seconds with a precision of better than 1 µm. A fast approach mechanism has been implemented on each miniaturized AFM instrument to engage the sample in less than 5 seconds. The miniaturized AFM instrument has a high-speed z-stage with a resonance frequency of 50 kHz and an optical readout with a bandwidth of 2 MHz and a noise of less than 15 fm/√Hz, enabling fast feedback and accurate tracking of surfaces even at high scan speeds. The high-speed z-stage is dynamically decoupled from the environment to eliminate disturbances due to high-frequency vibrations. Decoupling the PU from the miniaturized AFM instrument during scanning results in imaging of a large area, i.e., 10x10 µm$^2$, with minimal artifacts, noise and drift. The measured drift for a continuous scan of 27 minutes is 122 nm, or 0.075 nm per second. This demonstrator can open up new research opportunities in the nanometrology of wafers and lithography masks by enabling real die-to-die and wafer-level measurements and in cell biology by measuring the nano-scale properties of a large number of cells.

## ACKNOWLEDGMENTS

This research was supported by Netherlands Organization for Applied Scientific Research, TNO, Early Research Program 3D Nanomanufacturing Instruments. The research was performed under the framework of the E450EDL project. TNO gratefully acknowledges funding by the ECSEL Joint Undertaking and the Netherlands Enterprise Agency (RVO). The authors thank Dr. D. Piras for their helpful discussion on image analysis.

**REFERENCES**

[1]F. Ohnesorge and G. Binnig, Science **260,** 1451 (1993).

[2]R. A. Oliver, Rep. Prog. Phys. **71,** 076501 (2008).

[3]T. Ando, T. Uchihashi and T. Fukuma, Prog. Surf. Sci. **83,** 337 (2008).

[4]M. Kato, M. Ishibashi, S. Heike and T. Hashizume, Jpn. J. Appl. Phys. **41,** 4916 (2002).

[5]G. Binnig, C. F. Quate and C. Gerber, Phys. Rev. Lett. **56,** 930 (1986).

[6]U. Dürig, D. Pohl and F. Rohner, J. Appl. Phys. **59,** 3318 (1986).

[7]J. R. Matey and J. Blanc, J. Appl. Phys. **57,** 1437 (1985).

[8]A. Majumdar, Annu. Rev. Mater. Res. **29,** 505 (1999).

[9]Y. Martin and H. K. Wickramasinghe, Appl. Phys. Lett. **50,** 1455 (1987).

[10]H. Sadeghian, B. Dekker, R. Herfst, J. Winters, A. Eigenraam, R. Rijnbeek and N. Nulkes, Proceedings of in Proc. SPIE 9424, Metrology, Inspection, and Process Control for Microlithography XXIX, 94240O, San Jose, 2015.

[11]H. Sadeghian, N. B. Koster and T. C. van den Dool, Proceedings of in Proc. 2013 SPIE Advanced Lithography, 868127–868127–8, San Jose, 2013.

[12]M. Favre, J. Polesel-Maris, T. Overstolz, P. Niedermann, S. Dasen, G. Gruener, R. Ischer, P. Vettiger, M. Liley, H. Heinzelmann and A. Meister, J. Mol. Recognit. **24,** 446 (2011).

[13]T. Ando, T. Uchihashi and N. Kodera, Jpn. J. Appl. Phys. **51,** 08KA02 (2012).

[14]Y. Ahn, T. Ono and M. Esashi, J. Mech. Sci. Technol. **22,** 308 (2008).

[15]A. Schneider, R. H. Ibbotson, R. J. Dunn and E. Huq, Microelectron. Eng. **88,** 2390 (2011).

[16]S. C. Minne, S. R. Manalis and C. F. Quate, Appl. Phys. Lett. **67,** 3918 (1995).

[17]S. C. Minne, G. Yaralioglu, S. R. Manalis, J. D. Adams, J. Zesch, A. Atalar and C. F. Quate, Appl. Phys. Lett. **72,** 2340 (1998).

[18]M. Despont, J. Brugger, U. Drechsler, U. Dürig, W. Häberle, M. Lutwyche, H. Rothuizen, R. Stutz, R. Widmer, G. Binnig, H. Rohrer and P. Vettiger, Sens. Actuators A Phys. **80,** 100 (2000).

[19]H. Sadeghian, R. Herfst, J. Winters, W. Crowcombe, G. Kramer, T. van den Dool and M. H. van Es, Rev. Sci. Instrum. **86,** 113706 (2015).

[20]R. W. Herfst, W. A. Klop, M. Eschen, T. C. van den Dool, N. B. Koster and H. Sadeghian, Measurement **56,** 104 (2014).

[21]R. J. F. Bijster, J. de Vreugd and H. Sadeghian, Appl. Phys. Lett. **105,** 073109 (2014).

[22]R. Herfst, B. Dekker, G. Witvoet, W. Crowcombe, D. de Lange and H. Sadeghian, Rev. Sci. Instrum. **86,** 113703 (2015).

[23]H. Sadeghian, T. C. van den Dool, W. E. Crowcombe, R. W. Herfst, J. Winters, G. F. I. J. Kramer and N. B. Koster, Proceedings of in Proc. 2014 SPIE 9050, Metrology, Inspection, and Process Control for Microlithography XXVIII, 90501B, San Jose, California, United States, 2014.

[24]H. Sadeghian, R. W. Herfst, T. C. van den Dool, W. E. Crowcombe, J. Winters and G. F. I. J. Kramer., Proceedings of in Proc. SPIE 9231: 30th European Mask and Lithography Conference, Dresden, Germany, 2014.

[25]H. S. Marnani, T. C. Van Den Dool and N. Rijnveld, US Patent 9274138 (2016).

[26]AFM Gold Calibration Kit. Available: https://www.tedpella.com/calibration_html/AFM_Gold_Calibration_Kit.htm.

[27]Calibration Grating UMG01. Available: http://www.anfatec.com/cantilevers/umg01.html.

**FIGURES**

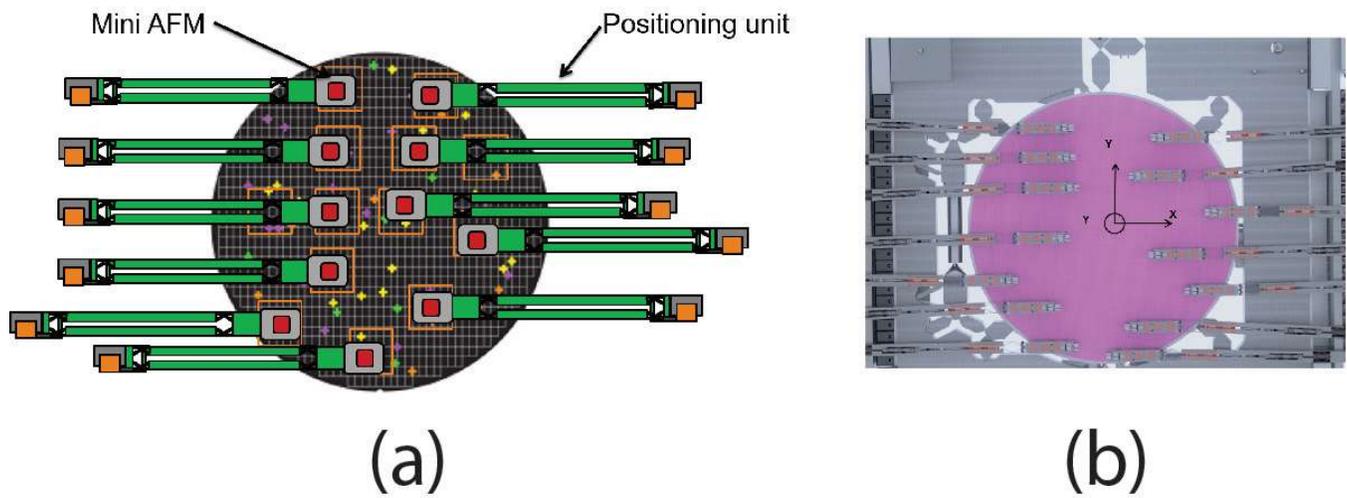

FIG. 1. (a) Schematic illustration of parallel AFM (bottom view) used to image several locations on a large area sample such as a wafer or a mask. (b) Computer-aided design of parallel AFM instrument showing the wafer stage in the middle and multiple positioning arms on two sides of the wafer stage, each capable of moving an MAFM instrument on to the wafer.

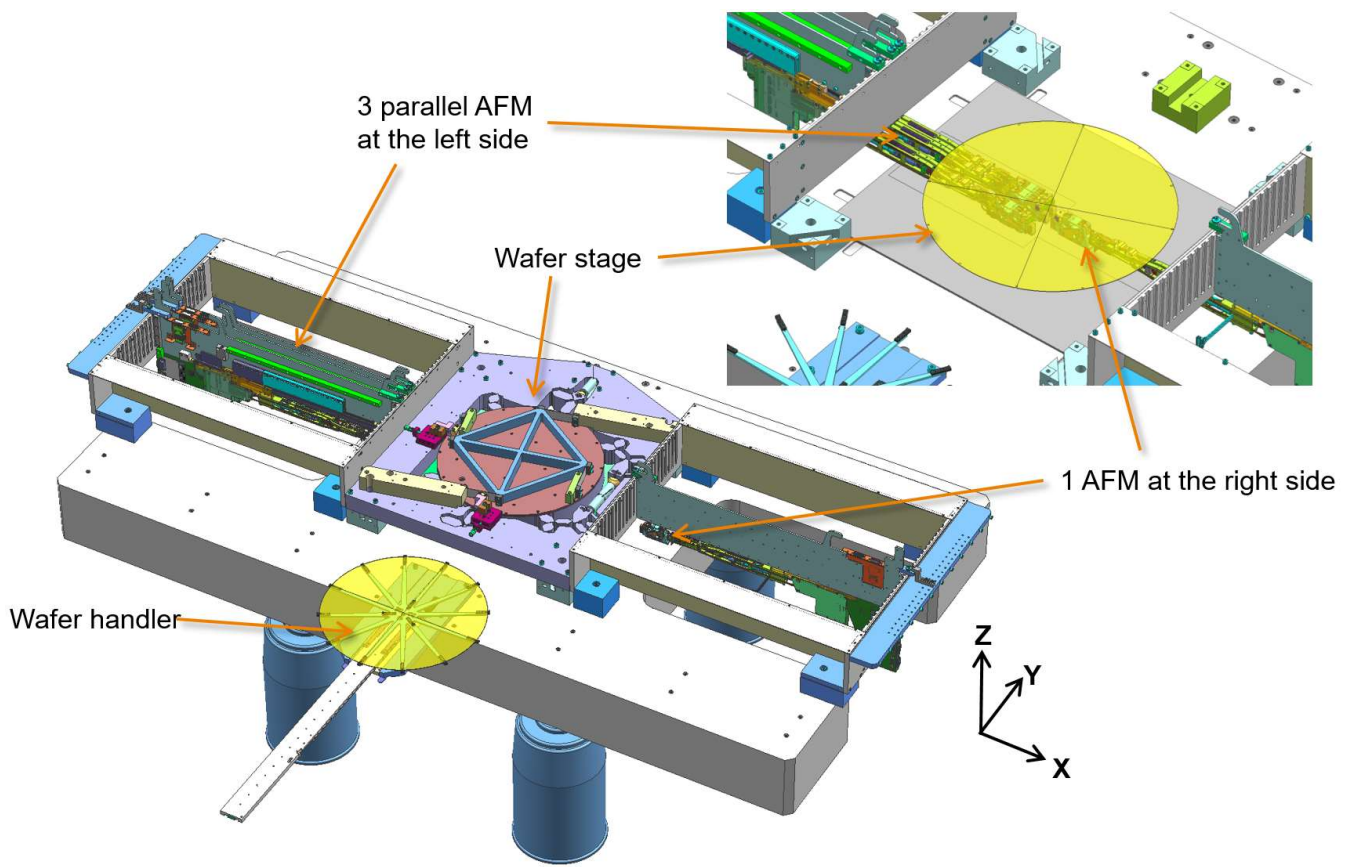

FIG. 2. CAD of the parallel AFM demonstrator. As seen, the wafer stage is in the middle, and the MAFM instruments are positioned toward the wafer using miniaturized PUs from two sides. The wafer is loaded onto the wafer stage with a manual wafer handler.

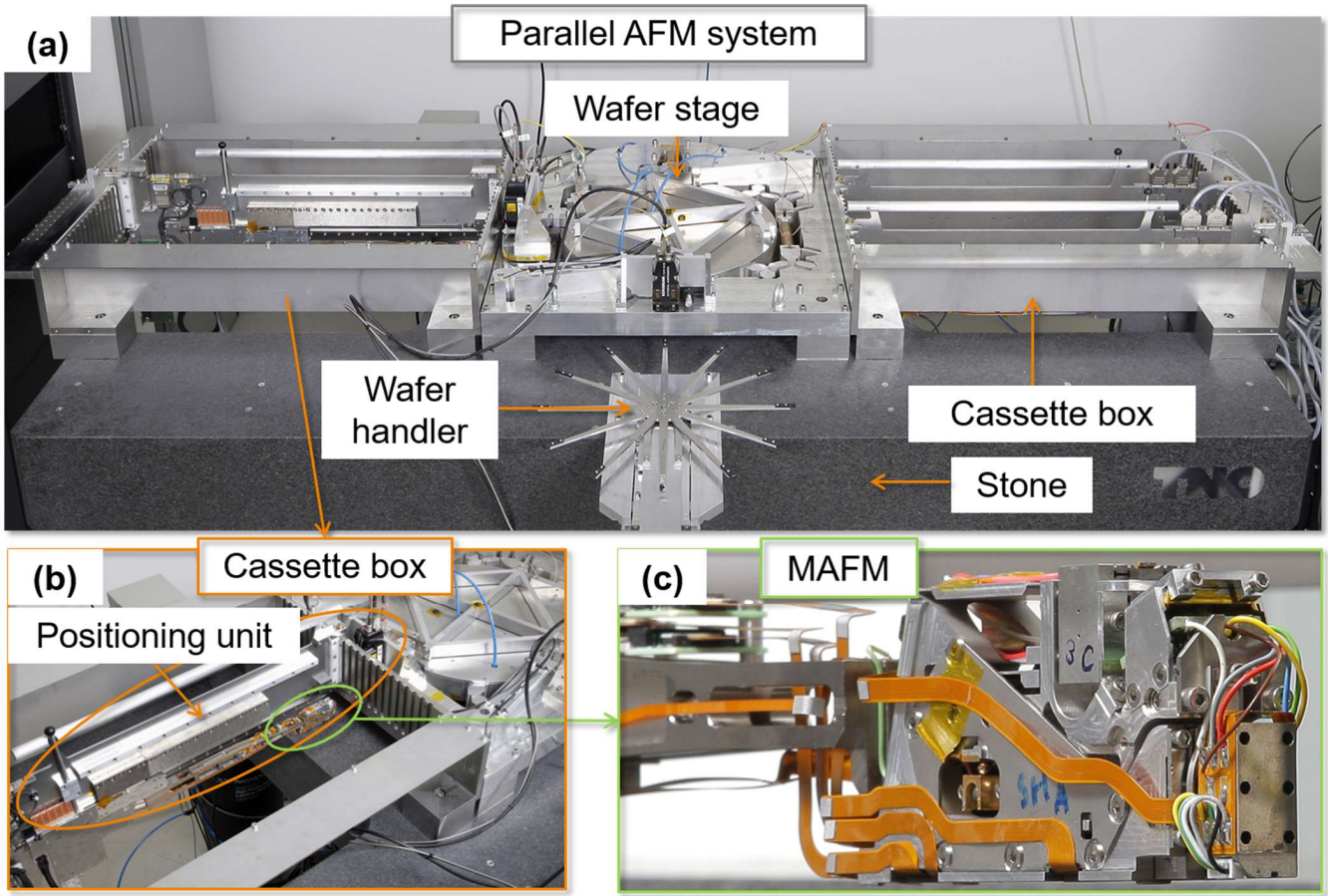

FIG. 3. (a) Parallel AFM demonstration. The wafer stage is shown in the middle. On the two sides, there are cassette boxes, each containing PUs and MAFM instruments. The wafer handler loads the wafer into the wafer stage. (b) One PU and one MAFM instrument are shown. (c) Two MAFM instruments are shown.

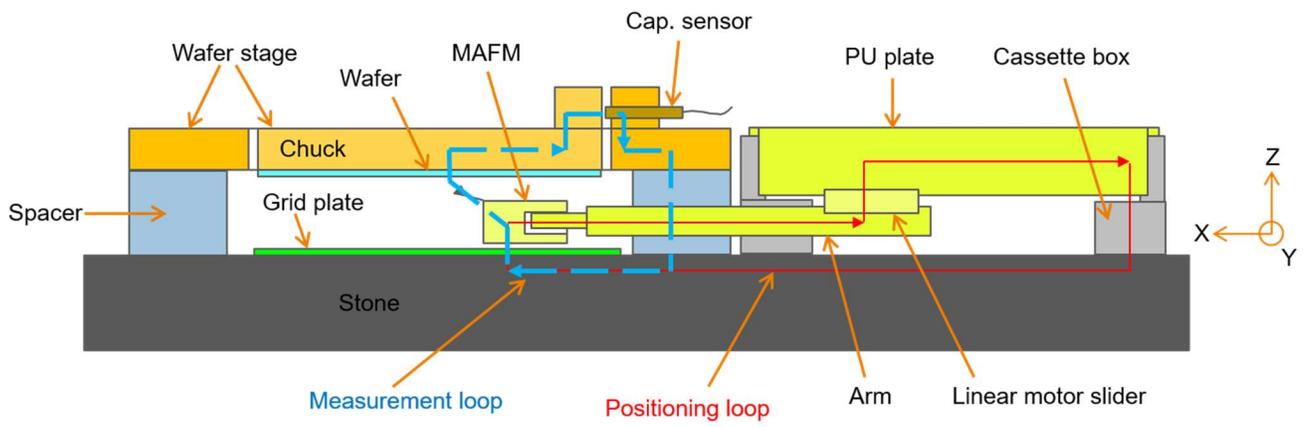

FIG. 4. Schematic overview of the components that constitute the positioning loop and the measurement loop.

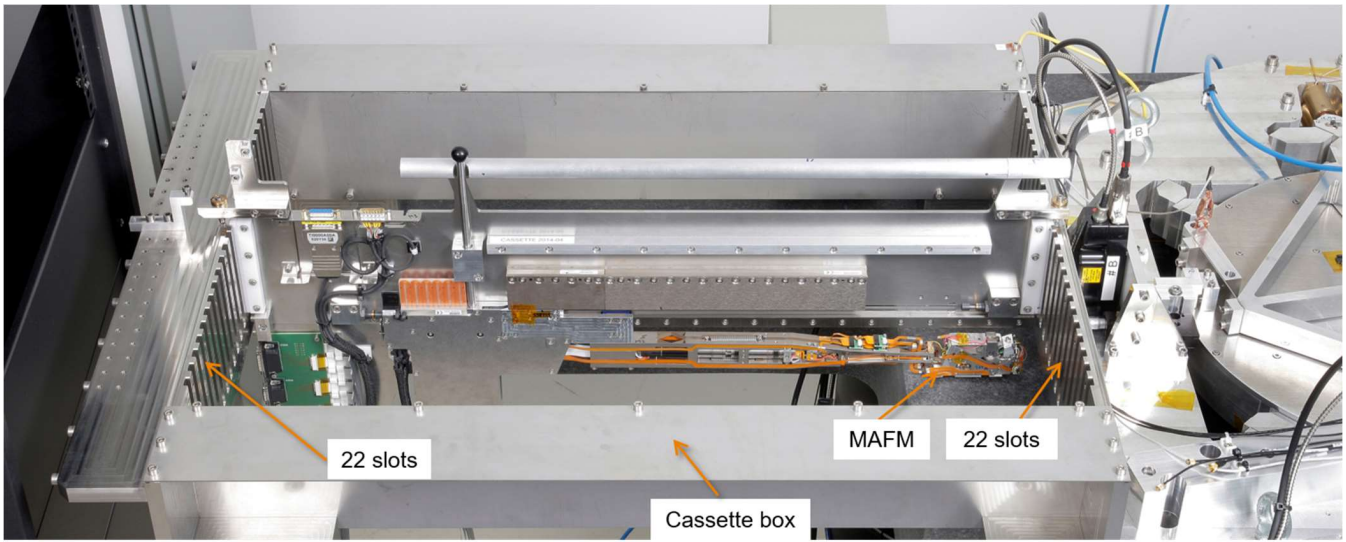

FIG. 5. The realized cassette box with one PU in place.

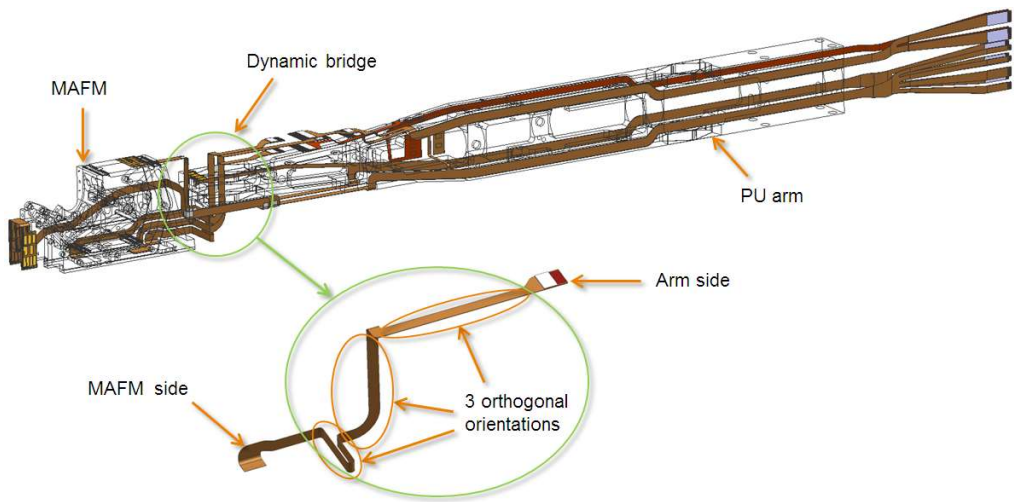

FIG. 6. Cable routing from the MAFM instrument to the arm, including the dynamic bridge crossover. The three orthogonal folding planes are shown for one of the flex-rigid cables.

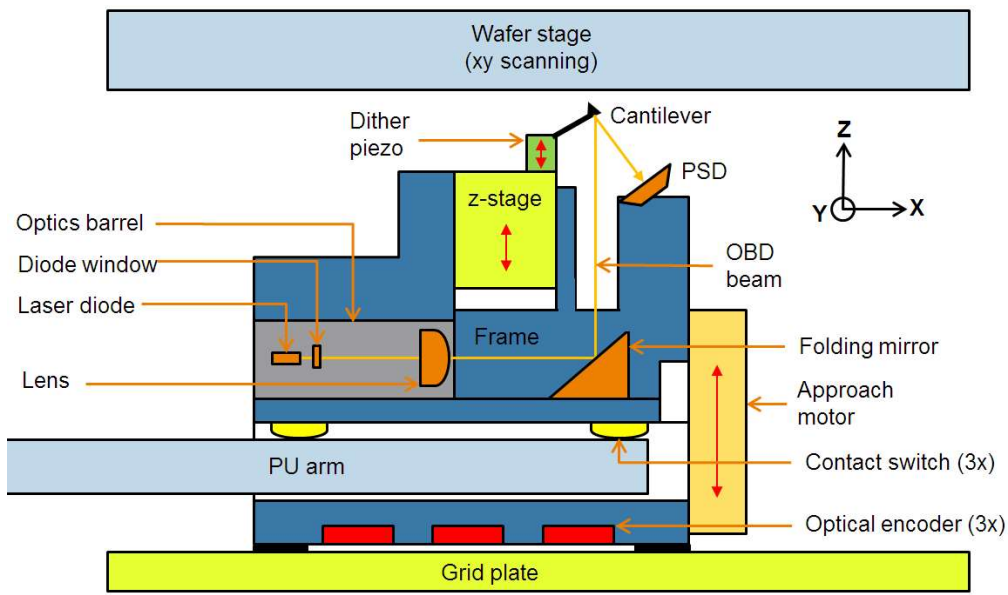

FIG. 7. Schematic overview of the MAFM instrument and its components.

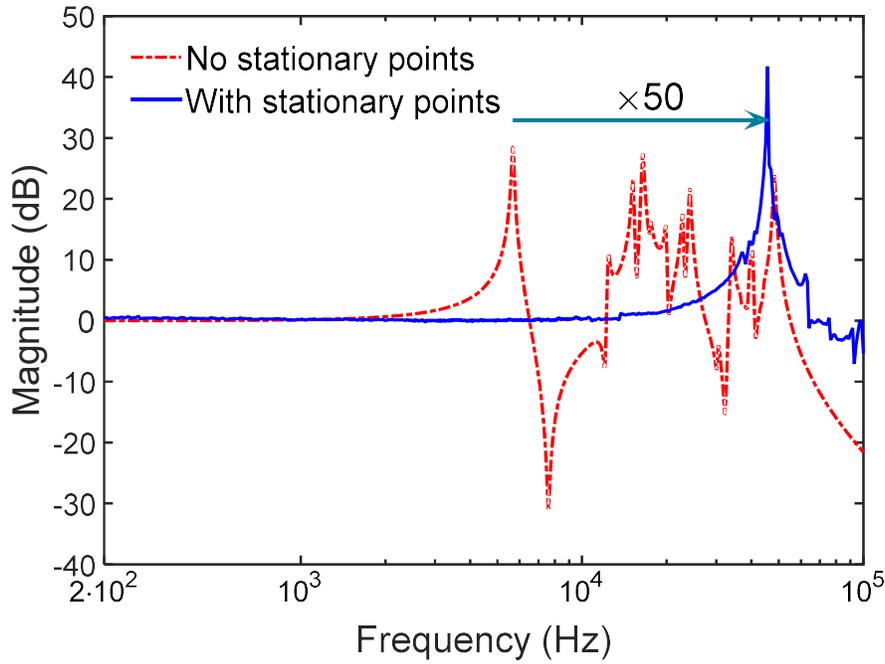

FIG. 8. Impact of balancing and using the stationary points for mounting the high-speed z-stage to the MAFM instrument frame. The dashed line is the expected frequency response without balancing and the use of stationary points. The solid line is the measured frequency response of the high-speed z-stage when the design principle of stationary points is implemented. Clearly, the first resonance mode has increased by a factor of 50.

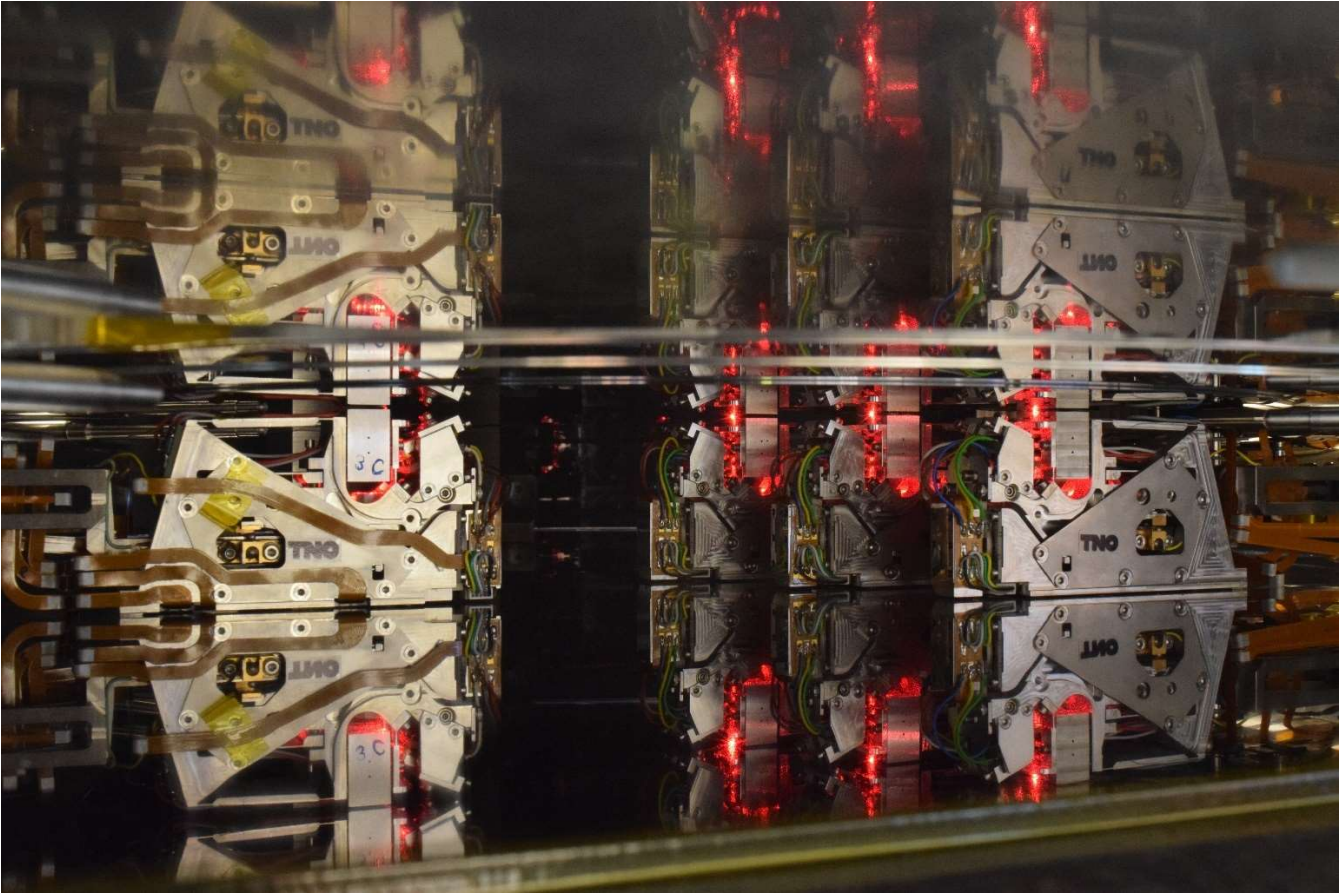

FIG. 9. Four MAFMs and their components.

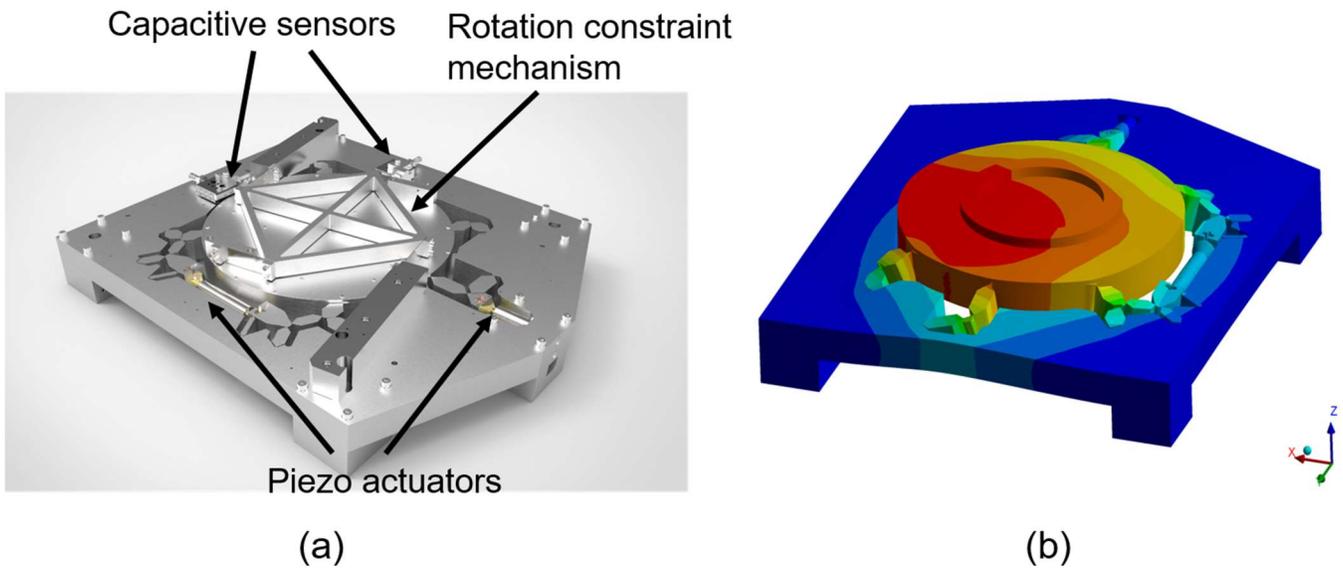

FIG. 10. (a) CAD of the wafer stage design showing the guidance and rotation constraint, capacitive sensors and the piezo actuation. (b) Modal analysis result of the wafer stage showing the mode shape of the lowest eigen frequency of 338 Hz.

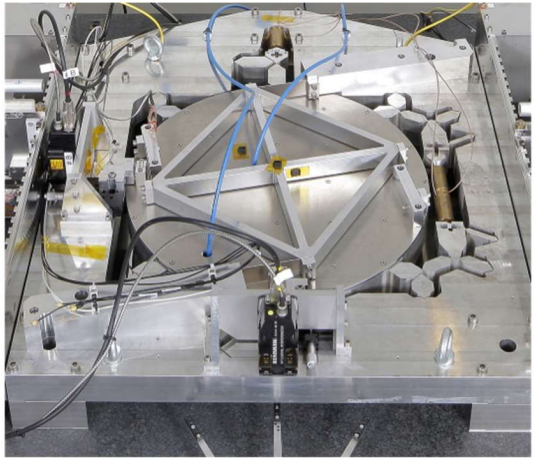

(a)

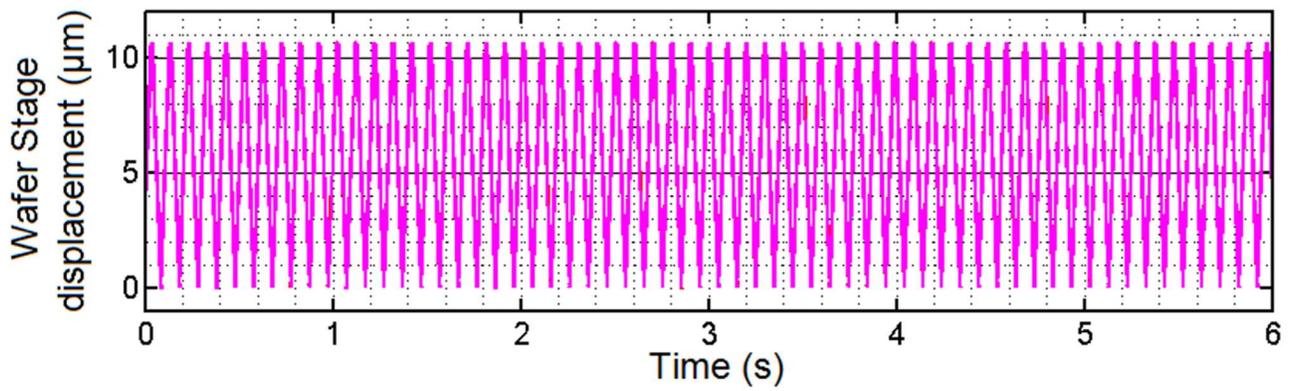

(b)

FIG. 11. (a) Manufactured wafer stage, showing the assembled stage, guidance and moving stage body, the capacitive sensors and the rotation constraint mechanism. (c) Measurement results of the wafer stage's motion speed, showing 20 lines/s at a 10.8 μm stroke.

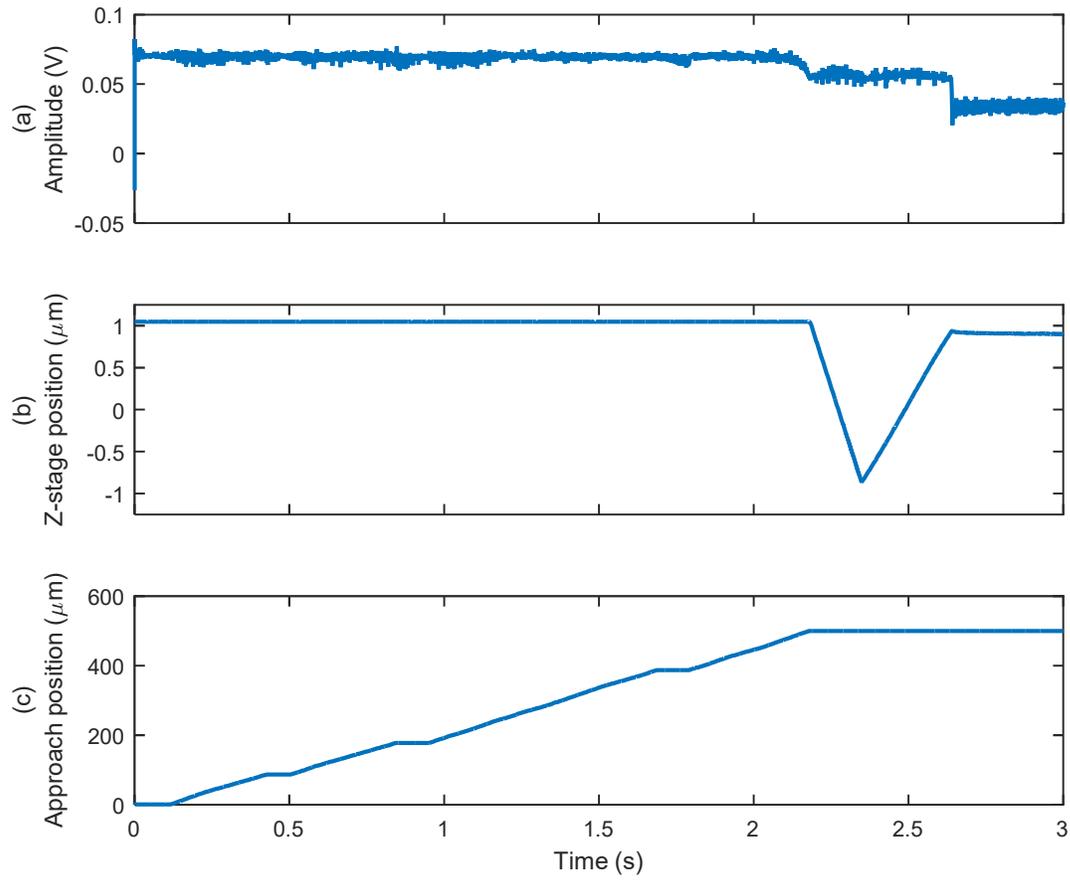

FIG. 12. Fast approach over 0.5 mm cantilever-sample distance, showing cantilever amplitude, fine z stage position, and coarse approach motor position as functions of time. At t = 0.11 s, the approach motor starts moving the cantilever toward the sample. Immediately before t = 2.18 s.

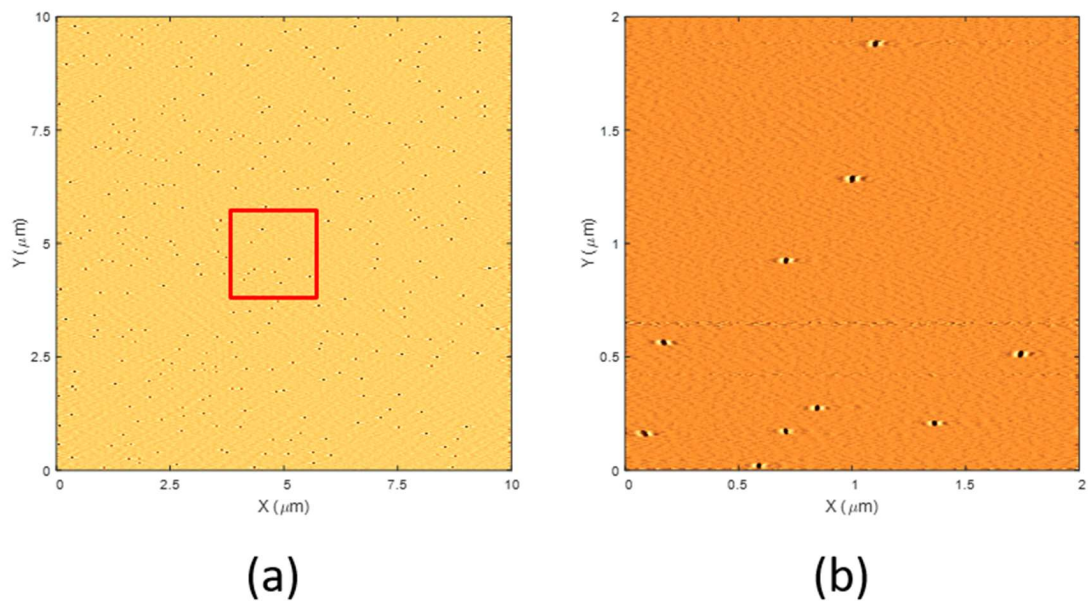

FIG. 13. AFM measurements of colloidal gold sample at 13 lines/s using an Arrow UHF cantilever at 1392 kHz. a) 10x10 μm, 2048x2048 pixels. b) Detailed measurement (2x2 μm, 2048x2048 pixels) of the selected section of (a).

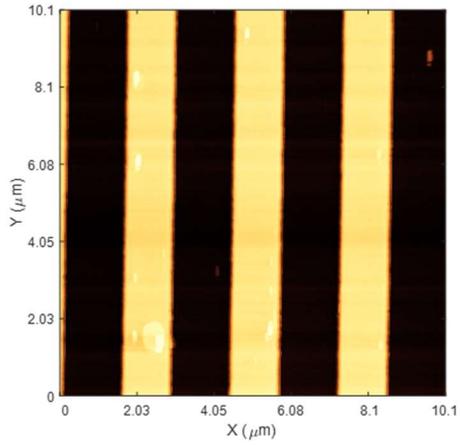
MAFM1

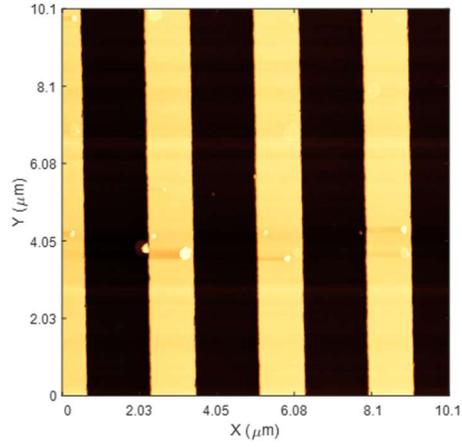
MAFM2

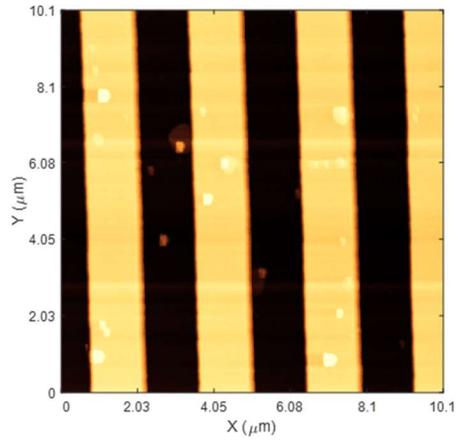
MAFM3

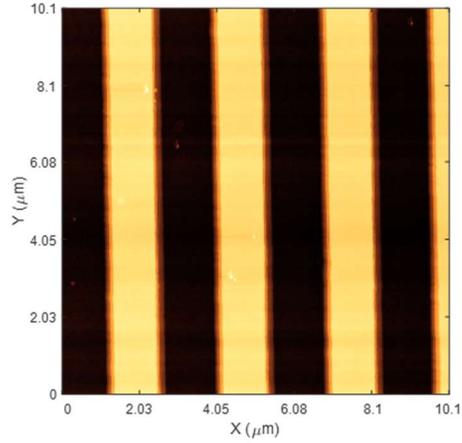
MAFM4

FIG. 14. Four simultaneously scanned 10×10 μm$^2$ images of line calibration samples with 2048×2048 pixels at a line rate of 2 Hz. All four MAFMs used an Arrow UHF cantilever.

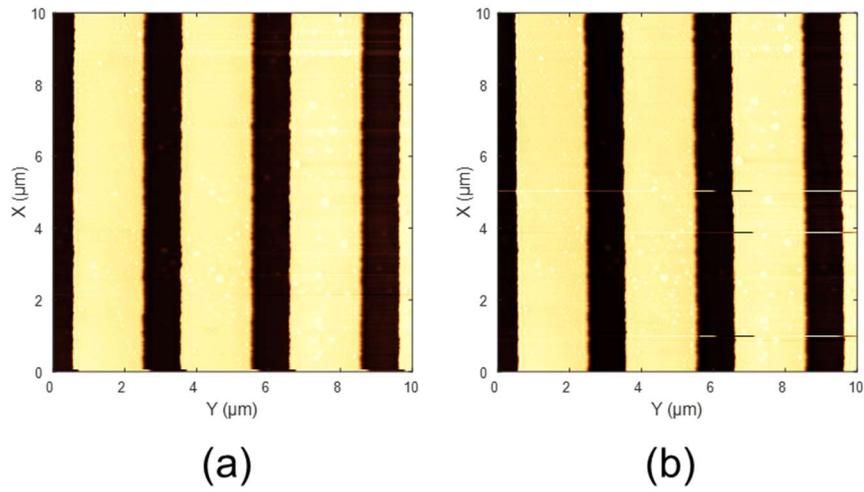

FIG. 15. Scanning image of the 3-μm-pitch, 110-nm-high line sample. (2048x2048 pixels, 2.5 Hz line-rate, Arrow UHF cantilever at 1144 kHz). a) Scan from top to bottom. b) Scan from bottom to top immediately after (a).

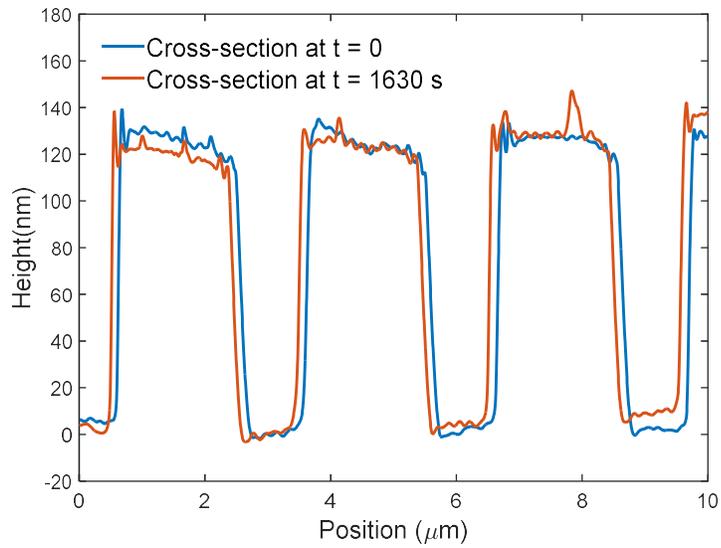

FIG. 16. Cross-sections of the line-sample scan of Figure 15 at the start and end of the drift measurement. A measurement time of 1630 seconds resulted in a 122 nm shift, i.e., a lateral drift of 0.075 nm/s.